\documentclass[prd,twocolumn,nofootinbib,showpacs,superscriptaddress]{revtex4}
\usepackage{amsmath}
\usepackage{epsfig}

\newcommand{\eqnref}[1]{Eq.~(\ref{eqn:#1})}
\newcommand{\figref}[1]{Fig.~\ref{fig:#1}}
\newcommand{\secref}[1]{Section~\ref{sec:#1}}
\newcommand{\appref}[1]{Appendix~\ref{app:#1}}

\newcommand{\fpder}[2]{\ensuremath{\frac{\partial #1}{\partial #2}}}
\newcommand{\fpderp}[2]{\ensuremath{\frac{\partial^{2} #1}{\partial #2^{2}}}}
\newcommand{\fpderthree}[2]{\ensuremath{\frac{\partial^{3} #1}{\partial #2^{3}}}}

\newcommand{\re}[1]{\ensuremath{\mathrm{Re}\{#1\}}}
\newcommand{\im}[1]{\ensuremath{\mathrm{Im}\{#1\}}}
\newcommand{\vect}[1]{\ensuremath{\vec{#1}}}

\begin{document}
\begin{abstract}
We present convergent gravitational waveforms extracted from 
three-dimensional, numerical simulations in the wave zone and
with causally disconnected boundaries.  These waveforms last for 
multiple periods and are very accurate, showing a peak error
to peak amplitude ratio of 2\% or better.  Our approach includes
defining the Weyl scalar $\Psi_4$ in terms of a three-plus-one
decomposition of the Einstein equations; applying, for the first time, 
a novel algorithm due to Misner for computing spherical harmonic
components of our wave data; and using fixed mesh refinement to 
focus resolution on non-linear sources while simultaneously resolving 
the wave zone
and maintaining a causally disconnected computational boundary.  We
apply our techniques to a (linear) Teukolsky wave, and then to
an equal mass, head-on collision of two black holes.  We argue both for the
quality of our results and for the value of these problems as standard
test cases for wave extraction techniques.
\end{abstract}

\pacs{04.25.Dm, 04.30.Nk, 04.30.Db}

\title{Wave zone extraction of gravitational radiation in three-dimensional
numerical relativity}
\author{David R.\ Fiske}
\altaffiliation[Current Address: ]{Decisive Analytics Corporation, 
1235 South Clark Street, Arlington, VA 22202}
\affiliation{Department of Physics, University of Maryland, College Park,
Maryland 20742-4111}
\affiliation{Laboratory for Gravitational Astrophysics, 
NASA Goddard Space Flight
Center, Greenbelt, Maryland 20771}
\author{John G.\ Baker}
\author{James R.\ \surname{van Meter}}
\affiliation{Laboratory for Gravitational Astrophysics, 
NASA Goddard Space Flight
Center, Greenbelt, Maryland 20771}
\author{Dae-Il Choi}
\affiliation{Universities Space Research Association, 10211 Wincopin Circle, 
Suite 620, Columbia, Maryland 21044}
\affiliation{Laboratory for Gravitational Astrophysics, 
NASA Goddard Space Flight
Center, Greenbelt, Maryland 20771}
\author{Joan M.\ Centrella}
\affiliation{Laboratory for Gravitational Astrophysics, 
NASA Goddard Space Flight
Center, Greenbelt, Maryland 20771}
\maketitle

\section{Introduction}
\label{sec:intro}
At the present time,
several ground-based gravitational wave detectors, using laser interferometry,
are already or very near to operating \cite{LIGO,LIGOII,schutz:sources},
and the NASA-ESA space-based antenna
LISA is scheduled to be launched in the next decade \cite{bender:lisa}.
These experiments should provide the first direct probe of strong-field
gravitational physics.  The data analysis needs of these observations, however,
require accurate waveform templates for use in matched-filtering
algorithms. While the early and late stages of a merger process can
be treated analytically using post-Newtonian and perturbation theory,
respectively, the highly dynamical merger period can only be understood
with the full, non-linear Einstein equations.  In this latter regime, numerical
relativity is essential \cite{schutz:sources}.

Computing waveforms from three-dimensional numerical simulations has, 
however, proven challenging 
\cite{camarda:waves,aei:distortedII,aei:grazing,aei:excision_dynamical}.  
One of the primary problems is that there are
a variety of length scales in a typical problem with sources.  For a
binary black hole collision, for example, the dynamics of the merger scale
with the masses and spins of the black holes, 
while the waves generated by those
dynamics have length scales that are one or two orders of magnitude larger.
In addition, the boundary conditions applied at the edges of the 
computational domain tend to generate spurious ingoing radiation
(which may or may not satisfy the full Einstein equations), which can
easily contaminate wave signals unless the boundary is causally disconnected
from the place and time at which the signal is desired.  
In addition, one must choose a stable form of the 3+1 Einstein
equations, supplemented by an appropriate set of gauge conditions, in order
to build a code (cf.\ Ref.~\cite{aei:gauge}).  Only recently has
it become possible to mix all of these components into a single simulation
in three dimensions. (See Refs.~\cite{aei:excision,maya:waves}
for very recent examples.)

In this paper, we bring together a variety of techniques and technologies
to successfully simulate dynamical spacetimes; to compute, via the Weyl scalar
$\Psi_4$, a gauge invariant measure of gravitational radiation; to analyze that
radiation quantity through spherical harmonic decomposition; and to demonstrate
that our approach not only converges very well, but that it also is
very accurate.  For our sample linear problem, a Teukolsky wave
propagating on a flat background, we see peak error to peak amplitude ratios
on the order of 0.4\%, and for our sample non-linear problem, the head-on
collision of two equal mass black holes, the same ratio is on the order of
2\% or less.  These waveforms maintain this level of accuracy for several
periods.

While we present our results, we also wish to lay out some well defined 
examples for future use as testbeds for wave extraction schemes, 
much in the spirit of the ``Apples with Apples''
tests for numerical relativity \cite{mexicoI}.\footnote{While this paper
was in final preparation, Ref.~\cite{maya:waves} was released as a preprint,
and, indeed, used very similiar test cases for wave extraction.
Ref.~\cite{maya:waves} focuses primarily, but not exclusively, on 
extractions via the Zerilli-Moncrief formalism rather than the Newman-Penrose
formalism that we use here.}
To this end, we attempt to
abstract away, where possible, from the details of our Einstein solver, and
focus on (1) the higher level details of how we define and compute
gravitational radiation quantities \emph{given} numerically evolved solutions
to the Einstein equations, (2) the description of a logical progression of 
test cases for validating results, and (3) the enumeration of
analytic solutions, symmetries, and cross-checks between test cases that lead,
when taken together,
to a high level of confidence in both the methods used and the results
obtained.

To these ends, we carefully
describe our particular approach to the problems of computing and 
analyzing gravitational waveforms in \secref{method}.  In that section, we
describe our method for computing the Weyl scalar $\Psi_4$ from our 3+1
numerical data, describe our particular algorithm for computing spherical
harmonic components of the Weyl scalar, and briefly describe our 
Einstein solver.  In \secref{waves} we bring this machinery to bear on the
Teukolsky wave spacetime.  This is a linear problem with an analytic
solution, which we use to validate our code and our methods.  We then
turn to a non-linear problem, the head-on collision of a binary black hole
system in \secref{headon}, which more closely resembles astrophysically
interesting sources and tests our methods on a non-linear problem.  We
conclude with some discussion of our results in \secref{discussion}.
The appendices contain more details about the spherical harmonic
decomposition algorithm and about the Teukolsky spacetime.

\section{Methodology}
\label{sec:method}
Computing gravitational waves from numerical simulations of the Einstein
equations requires combining a variety of nearly independent mathematical 
formalisms and
numerical techniques into a single code.   
The methods that we present, implement, and test here
are not specific to any particular formalism or to our code.  They are
made explicit in \secref{weyl}, where
we define gravitational radiation via the Weyl scalars 
in the Newman-Penrose formalism, and in \secref{Ylm}, where we
discuss the need and a method for computing spherical harmonic 
components of radiation data. In \secref{Hahndol}, 
we briefly discuss
our particular Einstein solver, with which we carry out numerical
simulations to validate our method. 

\subsection{Weyl Scalars}
\label{sec:weyl}
We use the Newman-Penrose formalism to compute gravitational radiation
quantities.  In this formalism, one chooses a tetrad of
four null basis vectors, which are 
conventionally labeled $l^a$, $n^a$, $m^a$, and 
$\bar{m}^a$. (Note that, although our code uses a
3+1 decomposition of the Einstein equations, indices on Newman-Penrose
quantities are always understood to run over four dimensional spacetime
indices $a = 0,1,2,3$.)
Of these vectors, $l^a$ and $n^a$ are real and point 
along ``outgoing'' and ``ingoing'' directions, respectively.
The vectors $m^a$ and $\bar{m}^a$ span 
``angular'' directions and are complex conjugates of each other.  These vectors
are always chosen to satisfy the orthogonality conditions
\begin{equation}
l^a m_a = n^a m_a = 0.
\end{equation}
In addition, we impose the normalization conditions
\(l^a n_a = -1\) and \(m^a \bar{m}_a = +1\),
which are used by most, but not all, authors (cf.\ Ref.~\cite{chandra}). 

Given a tetrad, tensor quantities can be recast as sets of coordinate scalars
by projecting tensor components onto the basis.  Applying this procedure to
the Weyl tensor\footnote{In vacuum, which is the only case that we consider 
here, the Weyl tensor is equal to the Riemann tensor.}
$C_{pqrs}$ yields five complex scalars.  This is particularly
useful for gravitational wave studies since one of the five,
\begin{equation}
\Psi_{4} = -C_{pqrs}n^{p}\bar{m}^{q}n^{r}\bar{m}^{s}, \label{eqn:Psi4}
\end{equation}
represents outgoing gravitational waves.  Indeed, the primary goals of this
paper are to construct $\Psi_4$ in some sample, numerically evolved spacetimes,
and to demonstrate that such computations are accurate characterizations of
the wave content of the spacetime.

Since, in our code, 
we have only 3+1 quantities on a single time slice
from which to compute the four dimensional Riemann tensor, we follow
Ref.~\cite{lazarus} and write
\begin{subequations}
\label{eqn:3dPsi4aux}
\begin{eqnarray}
^{(4)}R_{ijkl} & = & R_{ijkl} + 2K_{i[k}K_{l]j} \\
^{(4)}R_{0jkl} & = & -2\left( K_{j[k,l]} + \Gamma_{j[k}^{m}K_{l]m} \right) 
	\label{eqn:R0jkl}\\
^{(4)}R_{0j0l} & = & R_{jl} - K_{jm}K^{m}_{\ l} + KK_{jl}
\end{eqnarray}
\end{subequations}
which express four dimensional quantities on the left hand sides in terms of
3+1 quantities on the right.\footnote{There are two errors in 
Ref.~\cite{lazarus} associated with what we call \eqnref{3dPsi4aux}.  
First there is a factor of two difference between \eqnref{R0jkl}, 
which is correct, and the corresponding equation in
Ref.~\cite{lazarus}.  Second, \eqnref{3dPsi4aux} in this paper properly 
reflects the fact that the left hand
sides refer to the four dimensional Riemann tensor.}
Here the indices are spatial ($i,j,k,l,m=1,2,3$),
$K_{ij}$ is the extrinsic curvature tensor for a spatial slice as embedded
in the full spacetime manifold, and $K$ is its trace. 
The symbols $\Gamma^{m}_{jk}$, $R_{ijkl}$, and
$R_{jl}$ stand for, respectively, the connection coefficients, 
Riemann tensor, and Ricci tensor
associated with the three-dimensional spatial metric on the slice.

\subsection{Spherical harmonic decomposition}
\label{sec:Ylm}
In our simulations, we would like to be able to extract the spherical harmonic
components of gravitational waves.  We find this very valuable when analyzing
the data because
\begin{enumerate}
\item Numerical noise tends to have higher angular frequency 
than genuine wave signals, and is therefore filtered by the decomposition
process.
\item A priori knowledge about symmetries in the data or dominant modes
associated with physical processes allow important checks on the
plausibility of numerical solutions (especially when exact solutions are 
not available).
\item Some characteristics of gravitational radiation, such as 
quasi-normal modes for instance, are best understood in terms of 
spherical harmonic components.
\end{enumerate}

In practice, however,  we face a technical problem 
in computing spherical harmonic components of our data.
While our data is stored on a cubic grid, the spherical harmonic components
\begin{equation}
_{s}\Phi_{lm}(t,r) = 
\oint \mbox{}_{s}\bar{Y}_{lm}(\theta,\phi) \Phi(t,r,\theta,\phi) d\Omega
\label{eqn:sphdef}
\end{equation}
of a general (complex valued) function $\Phi$ in a spherical
harmonic basis with spin-weight $s$ are defined by an integral over a
sphere of radius $r$.  Computing the integral in \eqnref{sphdef} 
requires some type of
interpolation from grid points to points on that sphere.

One could solve this problem by the straightforward method of first
interpolating the grid function $\Phi$ to points on the sphere, and then
using some numerical approximation to the integral in 
\eqnref{sphdef}.  This process of interpolation 
followed by integration would
then need to be performed at each time, at each radius, and for every function
for which the spherical harmonic components are desired.

We adopt a different algorithm due to Misner.  Following 
Ref.~\cite{misner:Ylm}, we smear the surface integral in 
\eqnref{sphdef} into a volume integral
over a shell of half-thickness $\Delta$, and create an orthonormal basis
for functions on this shell by combining the spherical harmonics in the
angular directions with the first $N$ Legendre polynomials $P_n$ in the radial
direction.  This approach, which we describe in more 
detail in \appref{misner}, has the advantage that,
given the parameters $\Delta$ and $N$, one need only compute a relatively
small number of weights to carry out the volume integral, and that these
weights depend only on the grid structure and the radius of extraction.
This means that the weights can be computed once at the beginning of a 
simulation and stored.  Further interpolations are not needed, and the
weights are valid for all functions. (The weights do depend on the grid
structure, so they would needed to be recomputed after grid structure
changes in a simulation using adaptive mesh refinement.)

The question of how to choose the input parameters $\Delta$ and $N$ was
addressed in Refs.~\cite{fiske:phd,fiske:Misner_Ylm}.  Under the assumption
(motivated by the analysis in the references) that $\Delta$ is chosen
proportional to the grid spacing,
the parameter $N$ controls the order of convergence of the decomposition
algorithm, while the parameter $\Delta$ controls the size of the error at
that order.  This analysis leads to the 
\begin{quote}
\textbf{Rule of Thumb}: Choose $N$ just large enough to ensure that 
the error term proportional to
$\Delta$ is not of leading order in the grid spacing $h$.  Choose $\Delta$ just
large enough to safely resolve $P_N$ on the shell.
\end{quote}
For second order accurate codes such as ours, 
Refs.~\cite{fiske:phd,fiske:Misner_Ylm}
suggest choosing $N=2$ and $\Delta=3h/4$ based on this rule and empirical
experiments.  In cases where the
shell passes through multiple refinement regions, the grid spacing
of the \emph{coarsest} grid through which the shell passes should be chosen
to ensure that $P_N$ is resolved on both sides of the interface.  We follow
these suggestions for all work presented here.

\subsection{The Hahndol Code}
\label{sec:Hahndol}
Our code is a fully three-dimensional, non-linear Einstein solver.
We use a fairly standard formulation of the 3+1 Einstein evolution equations
known as BSSN \cite{sn:bssn,bs:bssn};
our particular implementation was described in detail in 
Ref.~\cite{goddard:puncture}.
Because the formulation of the equations is
not of primary interest here, and because the basics of the BSSN system
are widely known, we do not focus on these equations.

The code uses second-order accurate finite differencing to approximate
spatial derivatives and the iterative Crank-Nicholson method \cite{teuk:icn}
to integrate the evolution equations forward in time.  In both our previous
and current work, we have verified second order convergence in our results.

Our parallel code 
uses the \texttt{PARAMESH} libraries \cite{paramesh} to handle domain
decomposition and inter-processor synchronization.  In addition, these 
libraries enable us to use non-uniform grids to focus computational 
resources in specific areas of the computational domain.  Although the
libraries support, and we look forward to using, the ability to adaptively
modify the grid structure during the course of a simulation (adaptive
mesh refinement), we currently fix our grid structures in advance using
a priori estimates of where to focus resolution (fixed mesh refinement).
Our code shows 92\% of optimal scaling, up to 256 processors, when
simultaneously doubling the number of processors used for a simulation
and keeping the number of data points per processor 
constant \cite{goddard:puncture}.

Previous studies showed that the Hahndol code is able to propagate 
linearized 
gravitational waves (defined in terms of metric 
components using Teukolsky's solution \cite{teuk:wave}) 
across mesh refinement boundaries \cite{goddard:waves},
and that the same code can handle strong, dynamical potentials
moving across refinement regions \cite{goddard:puncture}.
In this paper we again focus on wave propagation, this time using a more 
formal definition of gravitational radiation, the Weyl scalars in the 
Newman-Penrose formalism, and on analyzing such data in a meaningful way.

\section{Teukolsky waves}
\label{sec:waves}
Our first problem studies Teukolsky's
solution \cite{teuk:wave} to the linearized Einstein equations, 
which represents
a weak gravitational wave propagating through space.  The linear nature of
the initial data makes this an excellent first test problem for two reasons.
First, the there is a analytic solution for all times, 
allowing a direct
calculation of numerical errors for convergence and accuracy studies.  Second,
because this solution consists of a linear wave in the initial data, 
we are able to extract our
waveforms at small radii and short evolution times.  This second fact is 
extremely useful for debugging algorithms, and also allows higher resolutions
for a fixed problem size since the volume of the computational domain
need not be as large as it would in problems (like the head-on collision
described in \secref{headon}) in which the waves are generated by non-linear
sources and must propagate to the wave zone before being extracted.

\subsection{Analytic Preliminaries}
\label{sec:teuk}

Although the Teukolsky solution is well-known, we summarize it here
for completeness and to establish notation.
The general form of the spacetime metric
\begin{eqnarray}
ds^2 & = & -dt^2 + (1 + Af_{rr})dr^2 \nonumber \\
& & \mbox{} + 2Bf_{r\theta}rdrd\theta 
+ 2Bf_{r\phi}r\sin\theta dr d\phi \nonumber \\ & &
\mbox{} + \left(1 + Cf_{\theta\theta}^{(1)} + Af_{\theta\theta}^{(2)}\right)
	r^{2}d\theta^2 \label{eqn:teuk} \\
& & \mbox{} + 2(A - 2C)f_{\theta\phi}r^{2}\sin\theta d\theta d\phi \nonumber \\
& & \mbox{} + \left(1 + Cf_{\phi\phi}^{(1)} 
+ Af_{\phi\phi}^{(2)}\right)r^{2}\sin^{2}\theta d\phi^2 \nonumber
\end{eqnarray}
is given in terms of angular functions $f_{ij}$ and functions $A$, $B$,
$C$.  The angular functions for the $l=2$, $m=0$ (spin-weight $-2$) 
case that we consider here are given by
Eqs.~(\ref{eqn:angular}) in \appref{teuk}.  The remaining functions, given by
Eqs.~(\ref{eqn:teukABC}), are written in terms of a 
free generating function $F$.

We follow Choi et al.\ \cite{goddard:waves} in choosing
\begin{equation}
F(x) = \frac{\mathcal{A}x}{\lambda^2} e^{-x^{2}/\lambda^{2}}
\label{eqn:F}
\end{equation}
as the exact form of the generating function, where the free parameters
$\mathcal{A}$ and $\lambda$ represent the amplitude and the width of the wave 
respectively.  The natural length unit in the problem is $\lambda$, and
we consistently choose \(\mathcal{A}=2 \times 10^{-6}\lambda\).
Moreover we take an equal superposition of an ingoing wave and an outgoing
wave, both centered at the origin, for the initial data.  This particular
choice has a moment of time symmetry that allows us to set the extrinsic
curvature tensor to zero in the initial data.

Choosing $F$ of the form in \eqnref{F} 
gives a waveform with oscillations but of 
essentially compact support.  This is ideal for testing codes with boundaries 
(both internal and external)
because it makes clear when the wave passes through those boundaries, and 
allows one to easily detect any reflections that occur due to poor
interface conditions; cf.\ Ref.~\cite{goddard:waves}.

For our 3+1 code, we use this analytic solution to generate the initial data,
and evolve forward in time using the gauge conditions lapse $\alpha=1$ and
shift $\beta^{i}=0$.

We use the Kinnersley tetrad \cite{kinnersley:tetrad}
\begin{subequations}
\label{eqn:Kinnersley}
\begin{eqnarray}
l & = & \frac{1}{\Delta}  (r^2+a^2,  \Delta, 0, a) \\
n & = & \frac{1}{2\Sigma} (r^2+a^2, -\Delta, 0, a) \\
m & = & \frac{1}{\sqrt{2}(r + ia\cos\theta)}
	(ia\sin\theta, 0, 1, i\csc\theta) \label{eqn:Kinnersley_m}
\end{eqnarray}
\end{subequations}
to extract the waves in this linear wave problem.\footnote{The first 
component of $m^{a}$ 
should have the $\sin\theta$ term in the numerator as in 
\eqnref{Kinnersley_m}; the corresponding equation in Ref.~\cite{lazarus}
is incorrect.}
Eqs.~(\ref{eqn:Kinnersley}) give the tetrad in terms of the Boyer-Lindquist
coordinates for a Kerr black hole. 
In these equations,
\(\Delta = r^{2} - 2Mr + a^{2}\),
\(\Sigma = r^{2} + a^{2}\cos^{2}\theta\), $M$ is the mass of the Kerr black 
hole, and $a$ is its spin.  Because Boyer-Lindquist coordinates are 
compatible with the coordinate system used for the Teukolsky metric
(\eqnref{teuk}) when $M=a=0$, we find this coordinate expression for 
the tetrad completely acceptable for this problem.

For this relatively simple spacetime, one can compute 
analytically the value of
the Newman-Penrose quantity
\begin{equation}
\Psi_4 = \frac{\sin^{2} \theta}{16} \left[
	-12 \fpderp{C}{t} + 6 \fpderp{A}{t} 
	+ r\left( 3\fpderthree{B}{t}
	+ \fpderthree{A}{t} \right) \right]
	\label{eqn:teuk_psi4},
\end{equation}
which represents outgoing gravitational radiation.  Noting that
\begin{subequations}
\begin{eqnarray}
_{-2}Y_{20}(\theta,\phi) & = & \sqrt{\frac{15}{32\pi}} \sin^{2}\theta \\
 & = & \sqrt{\frac{5}{6}} \left( Y_{00}(\theta,\phi) 
	- \frac{1}{\sqrt{5}} Y_{20}(\theta,\phi) \right)
\label{eqn:N2Y20}
\end{eqnarray}
\end{subequations}
it is clear that, as claimed, \eqnref{teuk_psi4} is a pure $l=2$,
$m=0$ mode.

Note also that \eqnref{N2Y20} provides a way to compute a
spin-weight $-2$ spherical harmonic component from the usual (spin-weight 0)
spherical harmonic components.
We make use of this in the results presented here because our current
implementation of the Misner algorithm computes only spin-weight 0
components, which suffices for our current work. In the near future we will
have the capability to compute spin-weight $-2$ components directly.

\subsection{Numerical Results}
\label{sec:numerics}
We evolved initial data specified by the Teukolsky solution described
in \secref{teuk} on a domain with outer boundaries at $32\lambda$ using
a mesh with fixed refinement levels having a 
``box-in-box'' structure centered on the origin.  The boundaries were 
at $2\lambda$, $4\lambda$, $8\lambda$, and $16\lambda$; neighboring regions
differed in resolution by a factor of two, with the finest regions 
surrounded by coarser regions.
We need to apply a matching condition at the interfaces between grids of
different resolutions; we use the algorithm described in
Ref.~\cite{goddard:puncture}.  We exploit the symmetry of the
solution to reduce our computational burden by evolving only one octant
of our data and using appropriate symmetry boundary conditions
to mimic a full grid.
In order to compute convergence factors, we ran the code at innermost
resolutions $\lambda/24$ and $\lambda/48$.  The wave
propagates outward from the origin, crossing the refinement boundaries
in turn.  We then attempted to compute the spherical harmonic
components of that wave at five distinct spherical ``detectors'' located at
$r=3\lambda$, $4\lambda$, $5\lambda$, $6\lambda$, and $7\lambda$.  

\begin{figure}[tb]
\begin{tabular}{c}
\epsfig{file=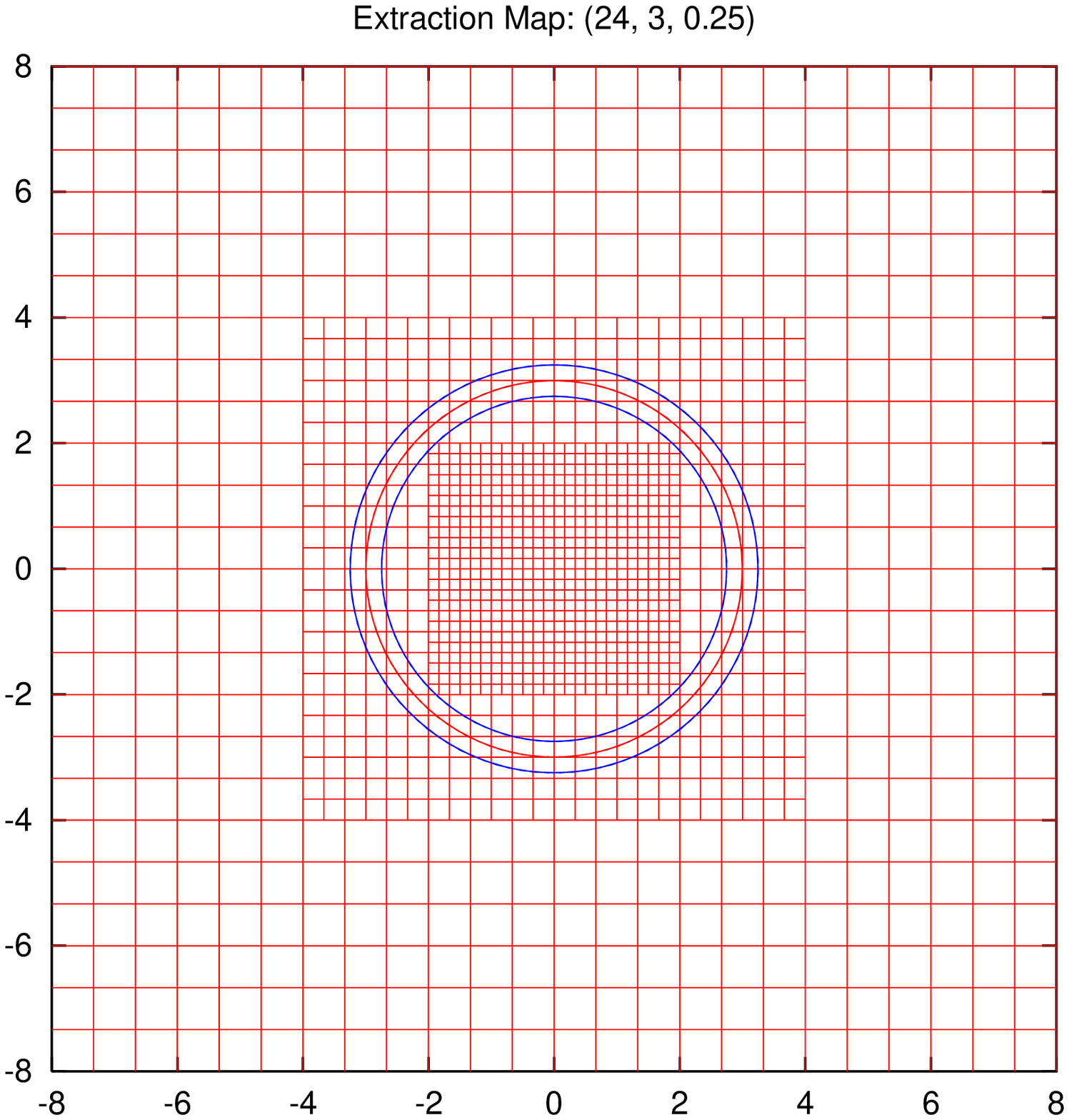,clip=true,width=2.4in} \\
\epsfig{file=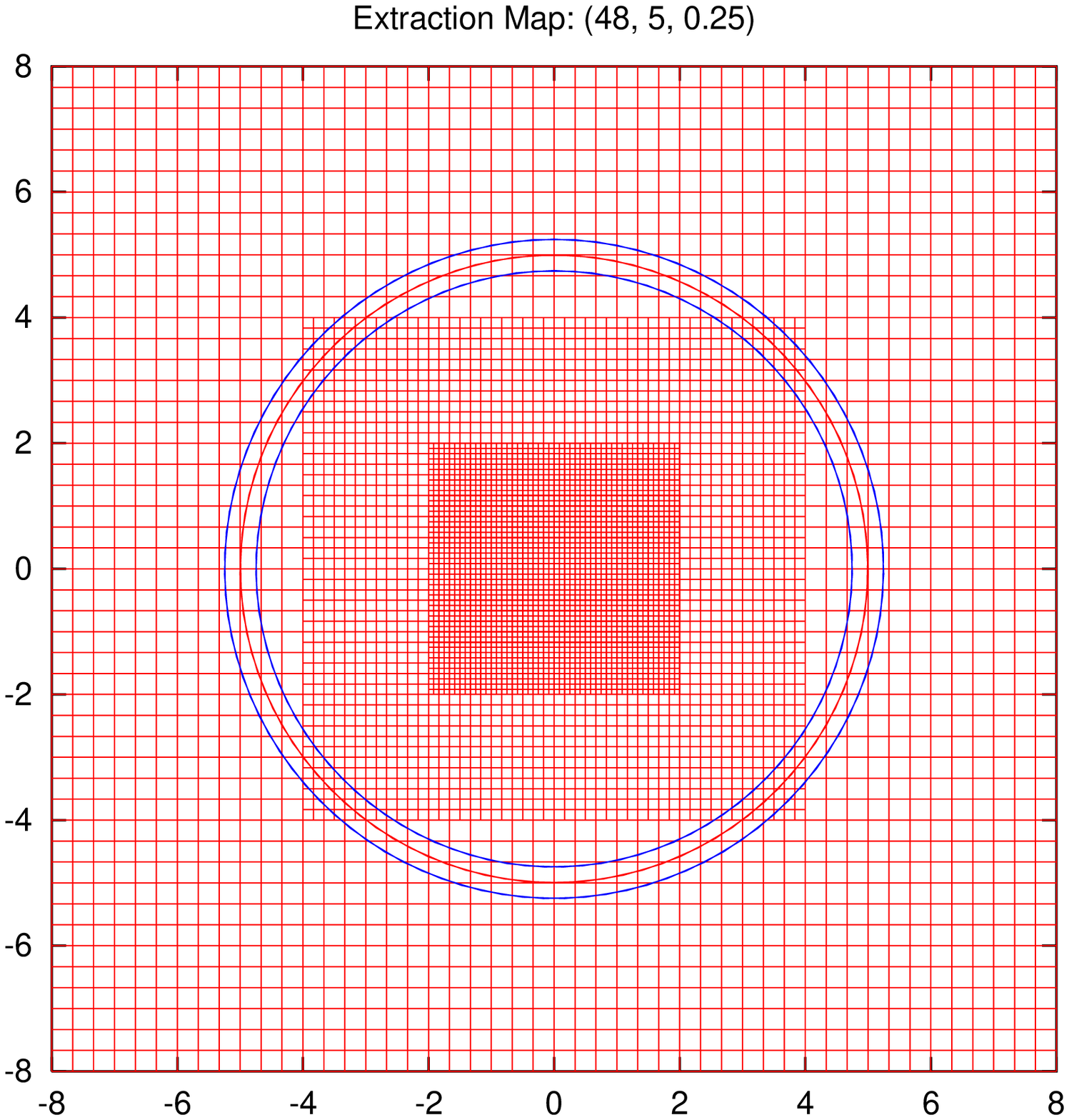,clip=true,width=2.4in}
\end{tabular}
\caption{Selected extraction maps.  Each map shows the computational
grid in a coordinate plane and is labeled with a triple of 
numbers $(N,R,\Delta)$, which indicate the number of (cell-centered) grid 
points across one coordinate direction in the coarsest region, 
the extraction radius, 
and the half-thickness of the shell, respectively.  Lengths are measured in
units of $\lambda$.
Note that the shells generically pass
through multiple refinement regions, especially since, in three dimensions,
the corners of the cubic refinement regions tend to poke through the
spherical extraction shells.  The three circles drawn on each graph show
the extraction sphere (red) and the edges of the finite-thickness shell
(blue) around the
sphere used in the Misner algorithm.}
\label{fig:extraction_maps}
\end{figure}

To help visualize how the refinement boundaries and the extraction radii 
are related geometrically, we find it useful to draw two
dimensional \emph{extraction maps}, as shown in \figref{extraction_maps}.
The extraction maps make it clear that the shells used to compute the
spherical harmonic components generically pass through multiple refinement
regions, especially since in three dimensions
the corners of the cubic refinement regions tend
to pass through the spherical extraction shells.  The extraction maps
shown in \figref{extraction_maps} correspond to the innermost ($r=3\lambda$)
extraction radius  and the $r=5\lambda$ extraction
radius in two sample resolutions.
(Extraction maps for
nearly all extraction radii and two resolutions appear in 
Ref.~\cite{fiske:phd}.) Note that only the innermost refinement regions
are shown in the maps.  In each case, the entire map was surrounded by
additional (coarser) refinement regions.

\begin{figure}[tb]
\begin{tabular}{c}
\epsfig{file=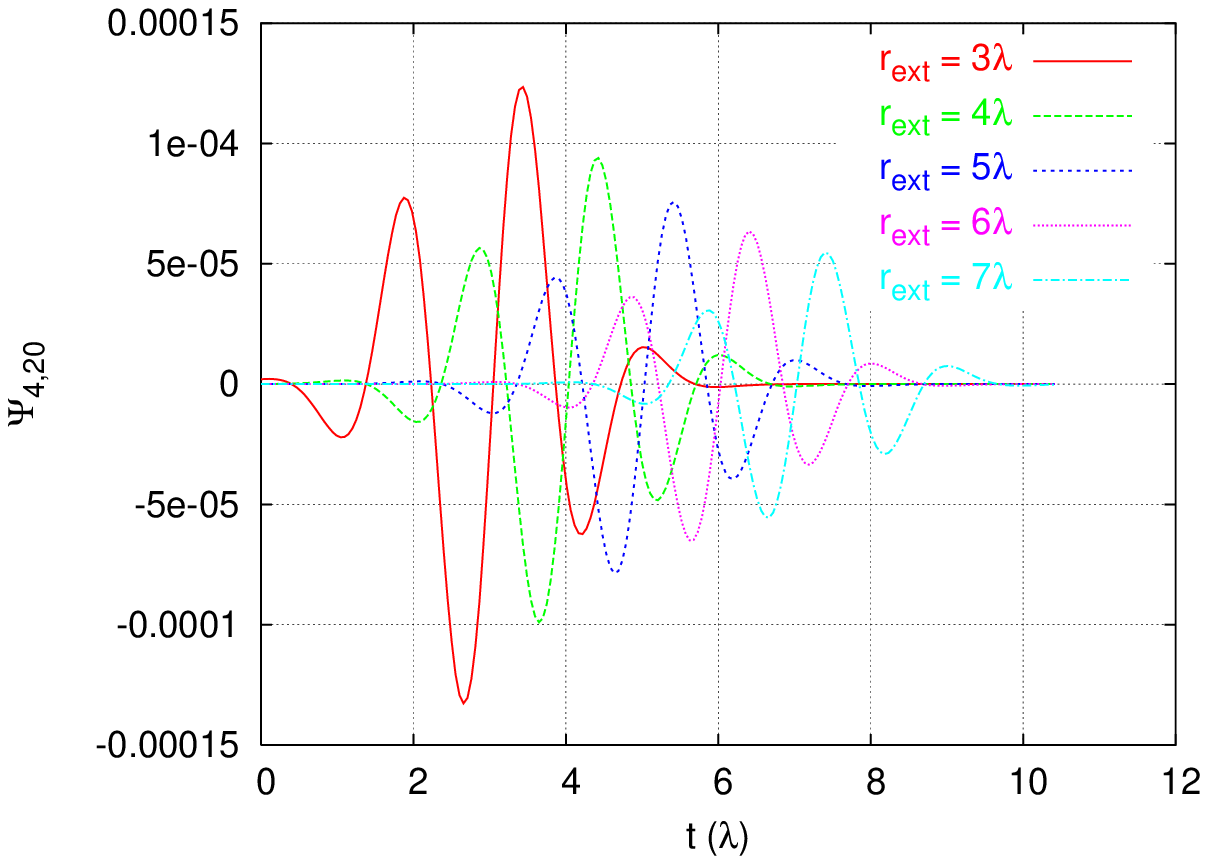,width=3.2in}  \\ (A) \\ 
\epsfig{file=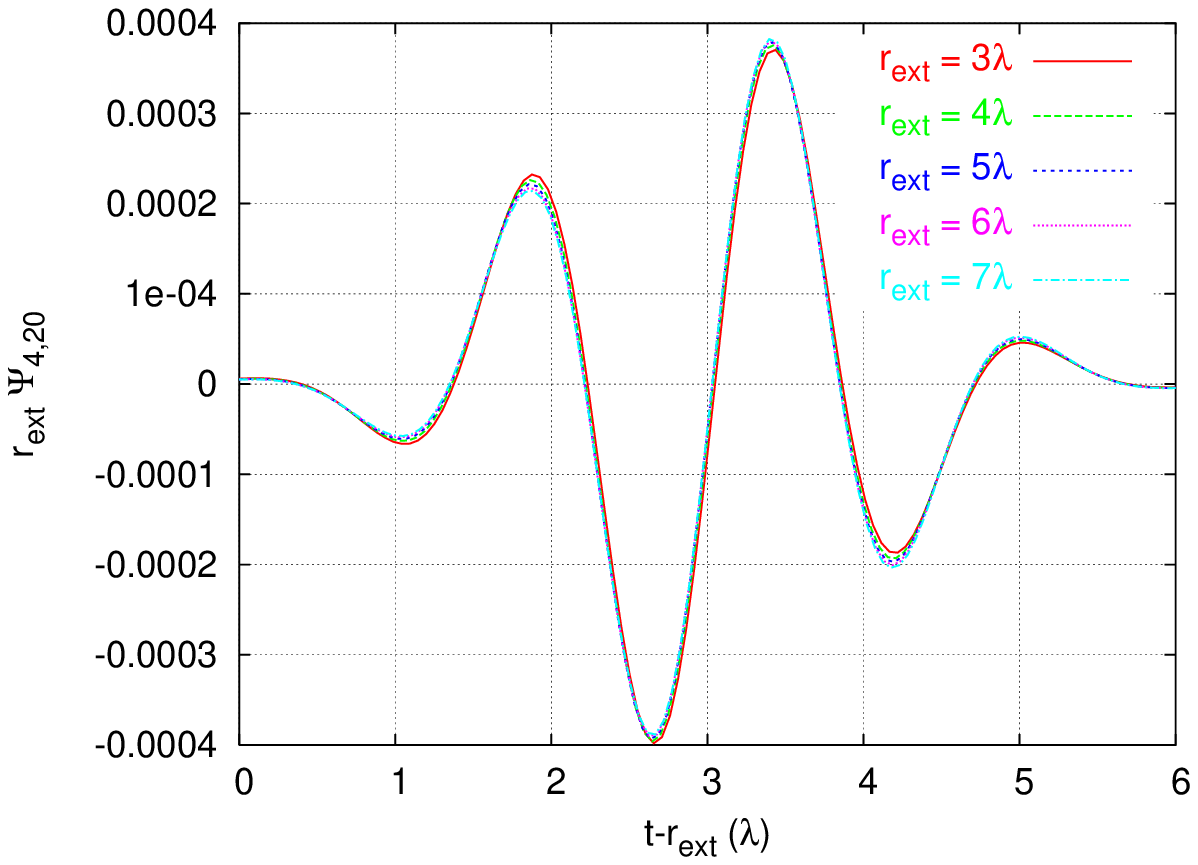,width=3.2in} \\ (B) 
\end{tabular}
\caption{The $l=2$, $m=0$ (spin-weight $-2$) component 
of the Teukolsky wave, as computed
at five different radii for the highest resolution run.  
The waveform is preserved up to the leading order $1/r$ scaling.
Panel~A shows the raw data. Panel~B shows the same data, scaled-up by 
$r_{ext}$ and shifted in time to $t=3\lambda$, the location of the inner-most
extraction radius.}
\label{fig:Psi4_teuk}
\end{figure}

We present our numerical waveforms in Panel~A of \figref{Psi4_teuk},
which shows the $l=2$, $m=0$ component of $\Psi_4$ for
a single wave as computed at our five distinct extraction radii.
The amplitude scales approximately as $1/r$, which is the correct behavior
to leading order in $1/r$, and the shape of the wave remains consistent as
it moves outward.  The consistency between the waveforms as extracted at
the various radii is demonstrated convincingly in Panel~B, in which we
have scaled-up the waveforms by a factor or $r$ and shifted them in 
time to $t=3\lambda$.  Assuming a unit speed of propagation, this should,
up to sub-leading terms in $1/r$, result in overlaying waveforms.  Indeed
the agreement demonstrated in Panel~B is striking.

In evaluating the effects of the refinement boundaries on the spherical 
harmonic decomposition algorithm, one should bear in mind that no two of
the extraction radii have the same geometric relationship with the
refinement boundaries (cf.\ \figref{extraction_maps}), and yet they nonetheless
generate perfectly consistent results.

\begin{figure}
\epsfig{file=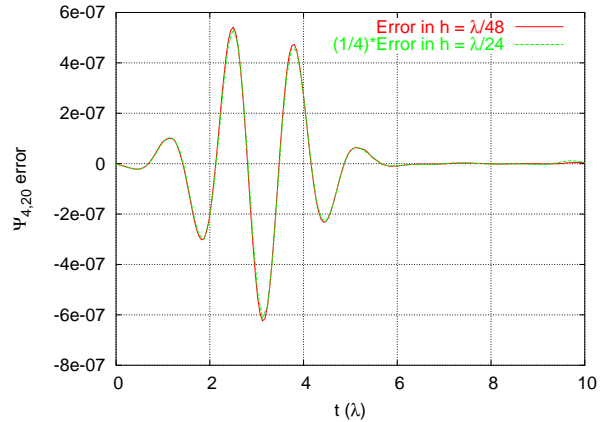,angle=-90,width=3.2in}
\caption{A convergence plot for the Teukolsky wave simulation.  The lines
are the errors in the $l=2$, $m=0$ (spin-weight $-2$) component of the 
simulations at the two resolutions as compared to the analytic 
solution.}
\label{fig:teuk_conv}
\end{figure}

Since we know the analytic solution for this test problem, we were able to make
a detailed convergence study with two resolution levels.  We verified
that not only the waveforms themselves but also the extracted spherical
harmonic components converge at second order, even with the
extraction spheres passing through the refinement boundaries, and
that the solution is highly accurate.
\figref{teuk_conv}
shows the errors, scaled by appropriate powers of two to indicate second-order
convergence, of the numerical results as compared to the analytic solution.  
It is clear from the figure that we do have second order convergence. Note from
\figref{teuk_conv} that our peak error in our highest resolution simulation is
approximately \(6 \times 10^{-7}\) for a waveform with peak amplitude of
approximately \(1.5 \times 10^{-4}\) (cf.\ Panel~A of \figref{Psi4_teuk}), 
giving a fractional error of approximately 0.4\%.

\section{Black hole head-on collision}
\label{sec:headon}

We also tested our techniques on a well-studied non-linear problem, 
the equal-mass, head-on collision of two black holes
\cite{aei:excision,lazarus:eclectic,lazarus:modeling,maya:waves}.
This problem is more closely related to the astrophysical problems of 
interest to us and, consistent with our
goal of testing our techniques, allows a variety of symmetries that we can
use as checks on our numerical solutions.

\subsection{Analytic Preliminaries}
\label{sec:headon_an}
In our head-on collisions, we place two black holes, each of mass $M/2$, on
the $z$ axis at a coordinate distance of $1.1515M$ from the origin.  The
initial data is generated by the puncture prescription of 
Ref.~\cite{brug:fmrexc}, which is a generalization of Brill-Lindquist
initial data \cite{brill-lind}.  We then use the excision algorithm of
Ref.~\cite{maya:moving} to avoid evolving the portions of our grid interior
to the (initially separate) apparent horizons of our initial data.  We neither 
allow our excision regions to move nor switch to a single, 
larger excision region
after a common horizon forms.  Although one or both of these techniques will
likely be desirable in future simulations, we find the simple approach
sufficient to extract convergent waveforms in this case.

Once we generate our initial data, we evolve it in our 3+1 code using
the 1+log evolution equation
\begin{equation}
\fpder{\alpha}{t} = -2\alpha K
\end{equation}
for the lapse $\alpha$, and the hyperbolic Gamma-driver evolution equations
\begin{subequations} \label{eqn:hGd}
\begin{eqnarray}
\fpder{\beta^i}{t} & = & \frac{3}{4}\alpha \psi_{BL}^{-4} B^i 
\label{eqn:hGd_beta} \\
\fpder{B^i}{t} & = & \fpder{\tilde{\Gamma}^i}{t}-\eta B^i 
\label{eqn:hGd_B}
\end{eqnarray}
\end{subequations}
for the shift $\beta^i$.  In these equations, $K$ is the trace of the extrinsic
curvature tensor for our slice, $\tilde{\Gamma}^i$ is the conformal 
connection function of the BSSN evolution system 
(cf.\ Ref.\ \cite{bs:bssn}), and $B^i$ is defined by \eqnref{hGd_beta}
to make the gauge evolution equations first order in time.  The quantity
\begin{equation}
\psi_{BL}(r) = 1 + \sum_{i=1}^{N} \frac{M_i}{2|\vect{r} - \vect{r}_i|}
\label{eqn:BL_factor}
\end{equation}
is the Brill-Lindquist factor used to generate puncture initial data for
$N$ black holes with masses $M_i$ and positions $\vect{r}_i$. 
(In our case, $N=2$ and $M_1=M_2=M/2$.)  In our work, we choose the 
parameter 
$\eta = 2.8/M$, and set $\alpha=1$ and $\beta^i=0$ at the initial time.  
These gauge conditions were first
studied in Ref.~\cite{aei:gauge}.

For the head-on collision we do not use the Kinnersley tetrad, since 
we have no way of knowing that
the coordinate expressions in \eqnref{Kinnersley} are appropriate to
this numerically evolved spacetime.  (Indeed, for a 
general problem, one would assume
that they are not appropriate.)  Instead, we
construct an orthonormal tetrad from the numerically evolved spacetime
using a Gram-Schmidt procedure as described 
in Ref.~\cite{lazarus}.\footnote{We used the Teukolsky wave as a check on our 
implementation of the Gram-Schmidt construction.  In the limit of 
perturbed flatspace, the Gram-Schmidt procedure recovers the Kinnersley 
tetrad up to a trivial rescaling of $l^a$ and $n^a$.  Comparisons of
waveforms extracted from Teukolsky wave spacetimes with the two choices 
of tetrad match very closely when this rescaling is taken into account.}

Since we do not have an analytic solution for the Weyl
scalars for the head-on case, we cannot directly compute the error 
in our solution. The symmetries of the problem nonetheless provide us with 
several analytic
checks on our numerical results.  Before discussing these symmetries
specifically, however, it is worth making explicit a related point:  There
are three independent symmetry axes in this (and any axisymmetic)
problem.  The first is the 
symmetry axis of the physical problem, i.e.\ the axis along which the two
black holes collide.  The second is the axis with respect to which the
tetrad is computed.  (Careful examination of the Kinnersley tetrad defined
by \eqnref{Kinnersley}, for example, reveals this explicitly in 
coordinate expressions.)  The third is the
axis with respect to which we compute the spherical harmonics.  Only if we
align all of these axes of symmetry (conventionally along the $z$ axis)
will all of our symmetry checks be true.\footnote{In addition, because,
for example, the inner product
\(\oint \mbox{}_{-2}\bar{Y}_{22}(\theta,\phi) 
\mbox{}_{0}{Y}_{l2}(\theta,\phi) d\Omega \neq 0\)
for all $l \geq 2$, it is impossible to construct the spin-weight
$-2$ components from spin-weight 0 components in problems lacking 
axial symmetry, unless one computes spherical harmonic components 
at all values of $l$ in the spin-weight 0 basis.}

Assuming that, as in the problem described here, all three symmetry axes
are aligned along the $z$-axis, the 
numerical solution should have the following properties:
\begin{itemize}
\item $\re{\Psi_4}$ is axially symmetric and is symmetric under the
transformation \(z \rightarrow -z\) up to the round-off level.
\item $\im{\Psi_4}$ converges to zero.
\item Viewed in a spin-weight $-2$ basis, the dominant contribution to
$\re{\Psi_4}$ comes from the $l=2$, $m=0$ mode.
\item Viewed in a spin-weight 0 basis, truncation level errors in 
$\im{\Psi_4}$ appear in only in modes with \emph{odd} values of $l \geq 3$
and \emph{even} values of $m \neq 0$.
\end{itemize}

\subsection{Numerical Results}
\label{sec:headon_num}
For the head-on collision simulations, we use a box-in-box
mesh refinement scheme similar to that employed in \secref{numerics}.
We place our outer boundary at coordinate distance
$128M$, and place refinement boundaries at $4M$, $8M$, $16M$, $32M$, 
and $64M$ (six levels total).
We find that in the non-zero shift case, we need to modify our matching
condition at the refinement boundaries; this will be the topic of
a forthcoming paper.  
We evolve only one octant of our spacetime, using
appropriate symmetry boundary conditions to mimic a full grid.
Because of the octant symmetry conditions used, only one of the black holes
appears in our evolved numerical grid.  (The other is accounted for by the
symmetry boundary conditions.)  We ran simulations at three resolutions with
innermost resolutions of $M/16$, $M/24$, and $M/32$ in order to perform 
convergence tests.

We excise a cubical region centered on each puncture.  The cube has
sides of length $0.23125M$.  The size, shape, and location of the excision
region remains fixed during the course of the run.

In these simulations, we extract our waveforms at $r=20M$, $30M$,
$40M$, and $50M$.  Since this is a non-linear problem, we are forced to use 
spheres at larger radii than in the Teukolsky wave simulations in order 
to ensure that the signals are being extracted in the wave zone.

\begin{figure}
\begin{tabular}{c}
\epsfig{file=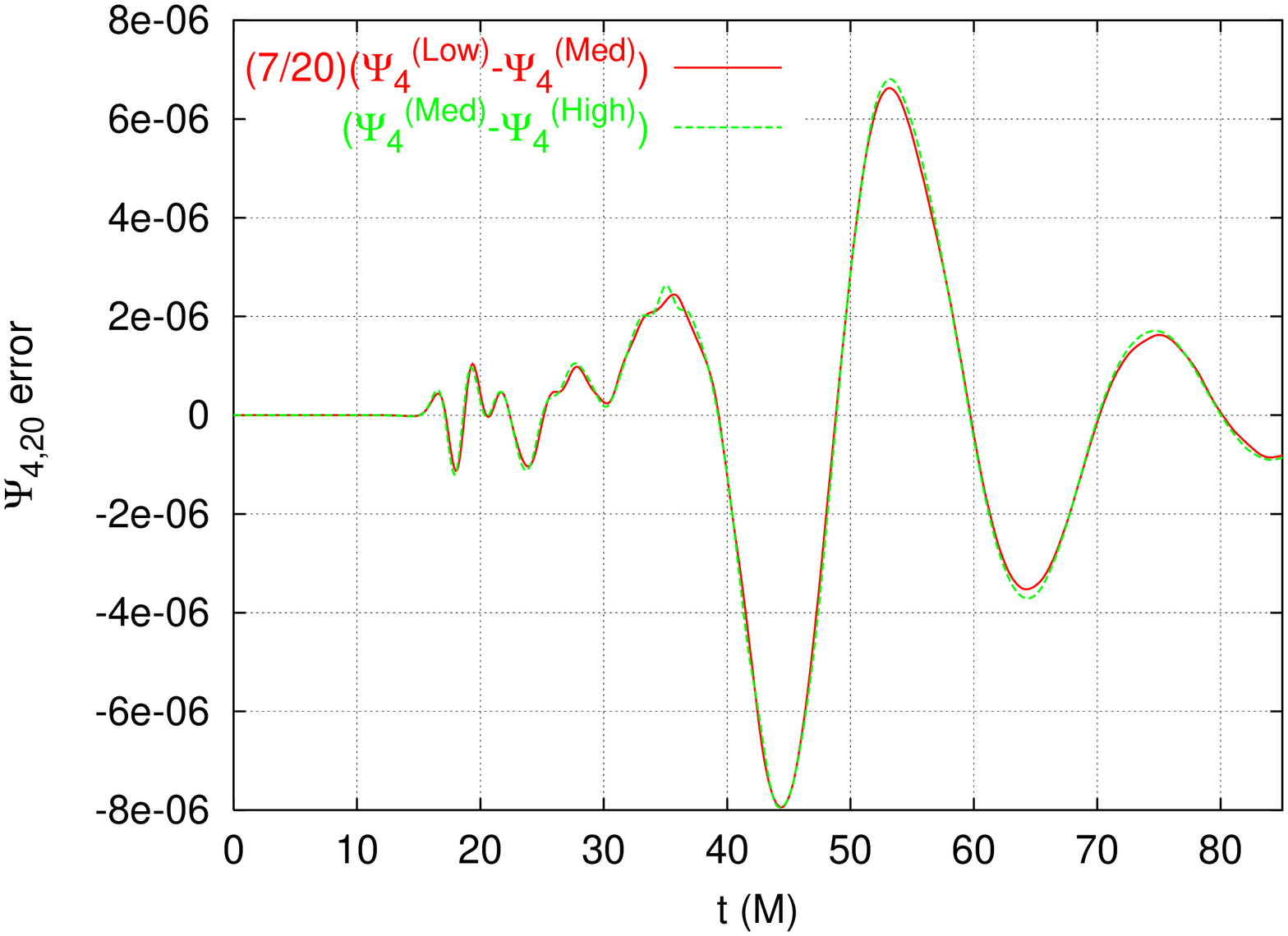,width=2.2in,angle=-90} \\ (A) \\
\epsfig{file=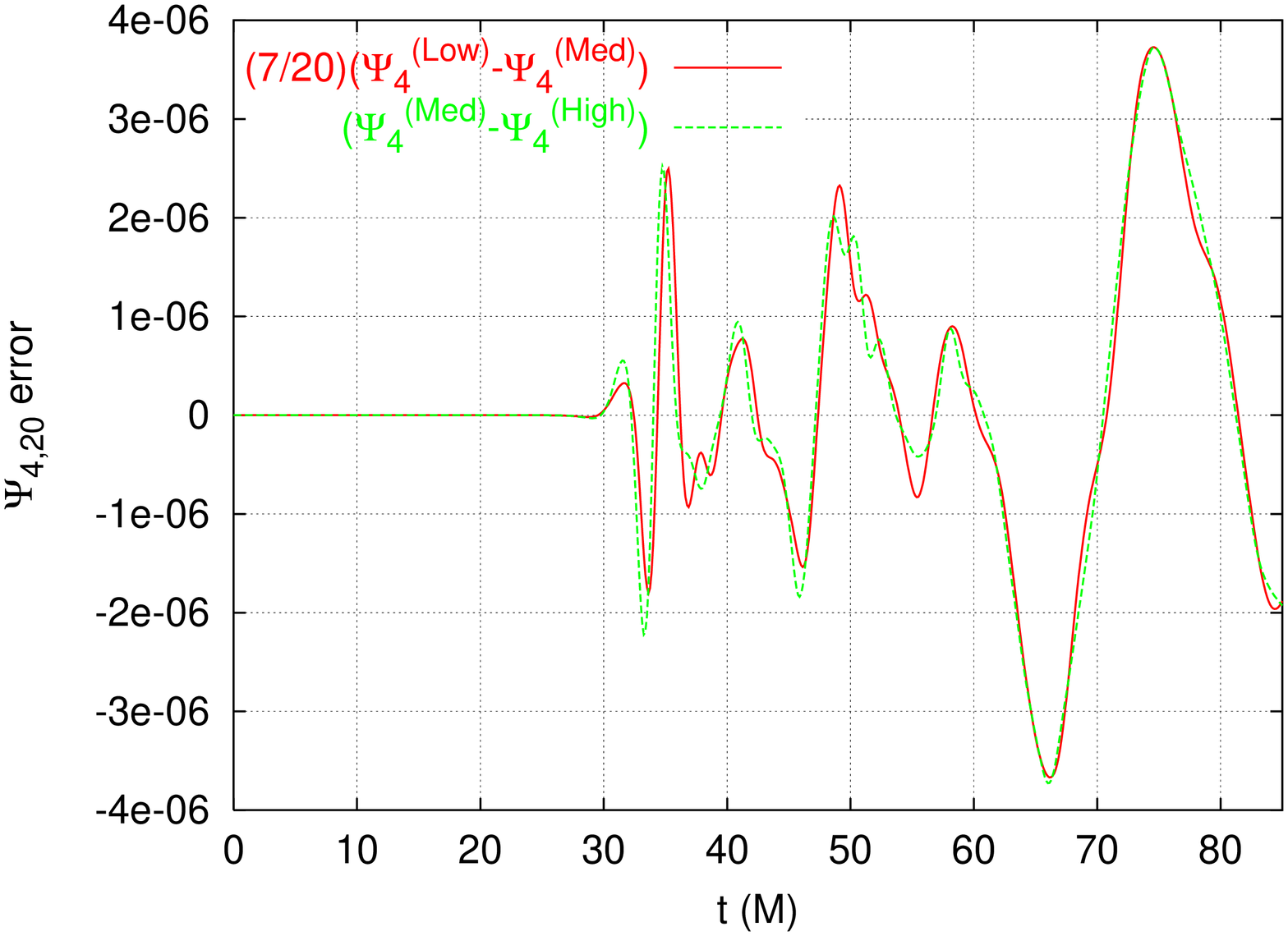,width=2.2in,angle=-90} \\ (B)
\end{tabular}
\caption{A convergence plot of the $l=2$, $m=0$ (spin-weight $-2$) component
of $\Psi_4$ for the equal mass black hole head-on collision problem.
Panel~A shows the convergence at the $r=20M$ extraction radius, and Panel~B
shows the convergence at the $r=40M$ extraction radius.  Note that, consistent
with the coarser resolution at larger radii, there is a slight decrease in
the agreement between the scaled errors in Panel~B, but that the agreement is
still excellent.}
\label{fig:head_convergence}
\end{figure}

Using the data from these three resolutions, we were able to evaluate the
convergence behavior of our waveform.  \figref{head_convergence}
shows a three-point convergence plot for the $l=2$, $m=0$ (spin-weight $-2$)
component of $\Psi_4$ extracted from our simulations.  In the three point
convergence graphs, we show
\((\Psi_{4}^{\mathrm{(Med)}} - \Psi_{4}^{\mathrm{(High)}})\)
and
\(7(\Psi_{4}^{\mathrm{(Low)}} - \Psi_{4}^{\mathrm{(Med)}})/20\),
which should coincide, up to the effect of higher order error terms,
for our second order accurate code, 
and our choice of relative grid spacings.\footnote{The unusual convergence 
factor 
\[\frac{7}{20} = \frac{(1/3)^2-(1/4)^2}{(1/2)^2-(1/3)^2}\]
is consistent with our simulations, which have resolutions
in the ratio \( 1/2:1/3:1/4\).}
The agreement is impressive.  Panel~A shows the convergence of the 
$l=2$, $m=0$ component of $\Psi_4$ as a function of time and as extracted
at radius $r=20M$.  Panel~B shows the same comparison for the extraction
radius at $r=40M$.  We observe that there is a \emph{slight} degradation of
the signal at larger radii, but that this is to be expected because the
larger radii are located in coarser regions of our grid.

\begin{figure}
\begin{tabular}{c}
\epsfig{file=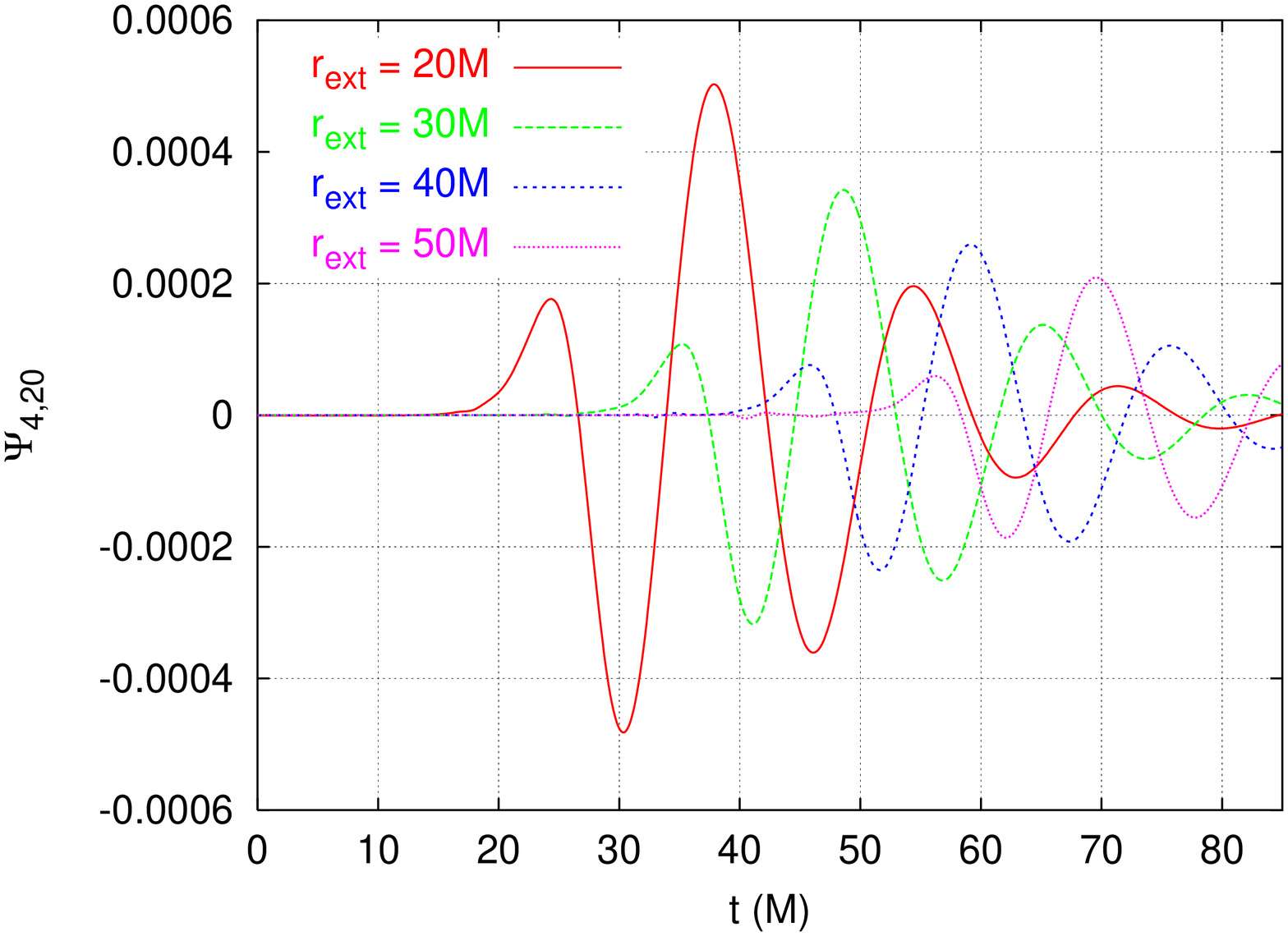,width=2.2in,angle=-90} \\ (A) \\
\epsfig{file=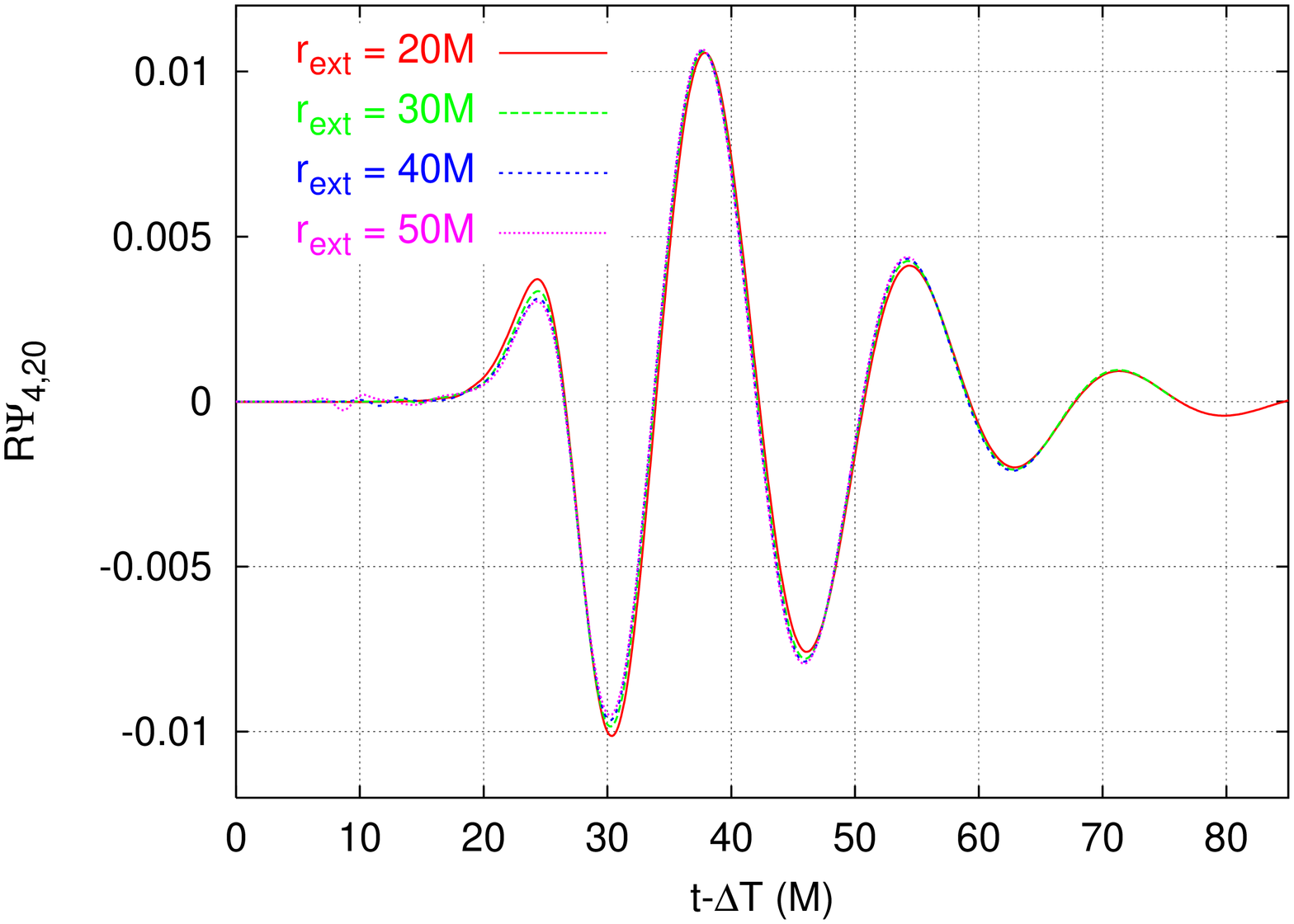,width=2.2in,angle=-90} \\ (B)
\end{tabular}
\caption{A comparison between the $l=2$, $m=0$ components of $\Psi_4$
as computed at distinct radii.  The waveforms have been scaled-up by $r$
and shifted to the innermost extraction radius, $r=20M$.  In order to 
account for the fact that our lapse is not geodesic, we scale
the waveforms by an approximate Schwarzschild radius \(R = r(1+M/(2r))^2\) 
rather than the simulation's radial coordinate $r$, and shift in time
by the light-travel time in a Schwarzschild background.}
\label{fig:head_scaled}
\end{figure}

In \figref{head_scaled}
we show the $l=2$, $m=0$ components of $\Psi_4$ as extracted at all four
radii.  As with the Teukolsky wave, we show, in Panel~A, the raw waveforms
plotted as a function of time, and, in Panel~B, the scaled and shifted 
waveforms.  Again note the excellent agreement between waveforms extracted
at different radii manifest in Panel~B.

It is important to note that our scaling and shifting 
procedure is more complicated in the head-on
collision case than it was in the Teukolsky wave case.  In the head-on 
case, we need to account for the fact that that the coordinate
speed of light is not spatially constant due to the presence of singularities.
We estimate the appropriate scaling factor and time shift by (1) assuming
that we extract our waves far enough from the binary system that the metric
is approximately Schwarzschild, (2) using the corresponding 
Schwarzschild radius
\begin{equation}
R = r \left( 1 + \frac{M}{2r} \right)^2
\label{eqn:Schwarz_R}
\end{equation}
as the scale factor rather that the simulation's coordinate radius $r$,
and (3) computing the time shift 
\begin{equation}
\Delta T = \left[ R' + 2M\ln(R'/2M-1) \right]_{R'=R_0}^{R}
\label{eqn:DeltaT}
\end{equation}
in terms of our reference (Schwarzschild) radius $R_0$.  The expression
in square brackets is often called the ``tortoise coordinate.''
In the case of \figref{head_scaled}, $R_0=R(20M)=21.0125M$.

\eqnref{Schwarz_R} was derived in Ref.~\cite{goddard:puncture} (among other
places), and is simply the coordinate conversion between the isotropic radial
coordinate $r$ and the Schwarzschild radius $R$ for a single, spinless
black hole.  It is, strictly speaking,
only applicable to our simulation on the initial time slice.  
At later times, our slices differ from the
Schwarzschild slices due to our choice of lapse $\alpha$; far enough from
the black holes, the slicings should be consistent and \eqnref{Schwarz_R}
should be a better and better approximation.

\eqnref{DeltaT} gives the Schwarzschild coordinate time $\Delta T$
required for an outgoing, radial, null ray to travel 
from radial coordinate $R_0$ to 
radial coordinate $R$.  It is also a good approximation to our coordinate time
difference $\Delta t$ in the limit of large radius.

Noting that both \eqnref{Schwarz_R} and \eqnref{DeltaT} are large-radius 
approximations to the appropriate scale factor and time shift, it is 
interesting to see that, although the waveforms extracted
at all radii shown in \figref{head_scaled} agree very well, the
extractions at the two farthest radii, $r=40M$ and $r=50M$, are extremely
consistent.  This is especially apparent in the first peak
at \(t-\Delta T \approx 25M\) and first trough at \(t-\Delta T \approx 30M\).

In addition to checking the convergence of our 
waveforms and verifying consistency between the waveforms computed at
different radii as shown above, we also verified that the four 
``sanity'' checks enumerated
at the end of \secref{headon_an} are satisfied, namely that both the real
and imaginary parts of $\Psi_4$ have all applicable symmetries, and that
the imaginary part converges to zero with increasing resolution.  
The combination of all of these checks gives us confidence in our results.

Since we do not have an analytic solution for this problem, we are not
able to directly compute the error in our simulation.  As a first estimate
of the accuracy that we achieve, we compute the ratio between the peak
error, as computed by the difference between the high and medium resolution
waveforms, and the peak amplitude of the high resolution waveform.  In the
worst case, our extraction at $r=50M$, this ratio is approximately 2\%.

\section{Discussion}
\label{sec:discussion}
We have presented a complete approach to computing gravitational radiation
via the Newman-Penrose formalism given a numerically evolved solution to
the 3+1 Einstein equations.  We validated our approach on the linear Teukolsky
wave problem, for which we have an analytic solution.  This allowed us to
very carefully study the convergence behavior of our results, and to
precisely measure the accuracy of the code.

Starting from the Teukolsky wave results, and combined with the 
technologies of mesh refinement and a spherical harmonic decomposition 
algorithm compatible with a non-uniform grid, we have also demonstrated that 
we are able to compute highly convergent waveforms generated by
genuinely non-linear sources at sufficiently far distances to be 
considered in the wave zone.  At the same time we locate the boundary 
of the computational domain at a sufficient distance to causally disconnect
it from the extraction region and still compute several cycles of the waveform.

The high quality of our waveforms, taken together with the consistency 
between waveforms extracted at various radii and with different 
geometric relationships to our mesh refinement boundaries, also validates
our particular choices for algorithms.  It specifically indicates that our
treatment of the mesh refinement boundary conditions does not introduce 
significant numerical errors into our simulation, and that our 
spherical harmonic decomposition algorithm, given good data on a non-uniform
grid, does not introduce any significant errors related to the relative
positions of the extraction radii and the refinement boundaries in the grid.

The work described in this paper sets the stage for us to study more
interesting astrophysical cases.  We are currently extending these
studies to non-equal mass collisions and inspirals, in which we will extend
our existing mechanisms to compute, based on the spherical
harmonic components of our waveforms, the energy and momentum
radiated by these systems.  We also hope that these results 
and these well-defined test 
cases can serve as a basis for future code validations and comparisons 
much in the spirit of Ref.~\cite{mexicoI}.

\begin{acknowledgments}
We would like to thank Charles Misner for providing advice as we implemented
his algorithm, and Richard Matzner who provided valuable guidance in the
development of these techniques.  We also thank Breno Imbiriba, who has 
contributed to the recent development of the Hahndol code.
We gratefully acknowledge the Commodity Cluster Computing Project
(NASA-GSFC) and Project Columbia (NASA Advanced Supercomputing Division,
NASA-Ames), which provided computing resources for the simulations
described here.
This work was supported in part by NASA Space Sciences grant ATP02-0043-0056.
DRF and JvM were also supported in part by the Research Associateship Programs 
Office of the National Research Council.
\end{acknowledgments}

\appendix
\section{Summary of Misner's Method}
\label{app:misner}
This appendix provides a more detailed look at the Misner algorithm for
computing spherical harmonic components of a function represented on a cubic
grid.  Additional details can be found in Misner's original paper, 
Ref.~\cite{misner:Ylm}.  A detailed discussion of the truncation error
as a function of the algorithm's parameters can be found in 
Refs.~\cite{fiske:phd,fiske:Misner_Ylm}.

Recall that the problem is to compute the spherical harmonic components 
$\Phi_{lm}$, defined by \eqnref{sphdef}, of a function $\Phi$ that is
known only on vertices of a cubic lattice. (In this appendix, we suppress
the spin-weight index $s$.)
In order to begin, two definitions are need.  First define a 
family of radial functions
\begin{equation}
R_{n}(r;R,\Delta) = r^{-1}\sqrt{\frac{2n+1}{2\Delta}}
P_{n}\left(\frac{r-R}{\Delta}\right)
\label{eqn:Rn}
\end{equation}
in terms of the usual Legendre polynomials $P_n$.  Here $R$ and $\Delta$ are
parameters that will be associated with the radius at which the spherical
harmonic decomposition is desired and half of the thickness of a shell 
centered on that radius.  From this, define
\begin{equation}
Y_{nlm}(r,\theta,\phi) = R_{n}(r)Y_{lm}(\theta,\phi)
\end{equation}
which form a complete, orthonormal set with respect to the inner product
\begin{equation}
\langle f | g \rangle = \int_{S} \bar{f}(x) g(x) d^{3}x
\label{eqn:continuumIP}
\end{equation}
on the shell
\( S = \{ (r,\theta,\phi) \mid r \in [R-\Delta,R+\Delta] \} \).
Note also that, because the functions $R_n$ form a complete set in the
radial direction,
\begin{equation}
\Phi_{lm}(t,R) = \int \rho(r;R,\Delta)
	\bar{Y}_{lm}(\theta,\phi) \Phi(r,\theta,\phi) d^{3}x
\label{eqn:PhiLMexact}
\end{equation}
with
\begin{equation}
\rho(r;R,\Delta) = 
\sum_{n=0}^{\infty} R_{n}(R;R,\Delta)R_{n}(r;R,\Delta),
\end{equation}
and that 
\begin{equation}
\rho(r;R,\Delta) = r^{-2}\delta(R-r)
\label{eqn:delta}
\end{equation}
is a delta function. (Compare \eqnref{PhiLMexact} 
to \eqnref{sphdef}.)

On a finite grid $\Gamma$, the inner product \eqnref{continuumIP}
will have the form
\begin{equation}
\langle f | g \rangle = \sum_{x\in\Gamma} \bar{f}(x) g(x) w_{x}
\label{eqn:numip}
\end{equation}
where each point has some weight $w_{x}$.  This weight was given the
form
\begin{equation}
w_{x} = \left\{ 
\begin{array}{ll}
0   & |r-R| > \Delta + h/2 \\
h^3 & |r-R| < \Delta - h/2 \\
(\Delta +h/2 - |R-r|)h^2 & \mbox{otherwise}
\end{array}\right.
\label{eqn:MisnerWeights}
\end{equation}
by Misner, where $h$ is the grid spacing.  Only cases with
$\Delta > h/2$ are considered.
This means, roughly, that points entirely within the 
shell $S$ are weighted by their finite volume on the numerical grid,
points entirely outside of the shell $S$ have zero weight, and points near
the boundary are weighted according to the fraction of their volume 
inside $S$.

With the numerical inner product \eqnref{numip}, and letting capital
Roman letters \(A = (nlm)\) represent index groups, the $Y_{A}$ are no
longer orthonormal.  Their inner product
\begin{equation}
\langle Y_A | Y_B \rangle = G_{AB} = \bar{G}_{BA}
\label{eqn:YlmMetric}
\end{equation}
forms a metric for functions on the shell. (Although a priori 
this matrix appears to be complex valued, it is
actually real-symmetric and sparse, 
cf.\ Refs.\cite{fiske:phd,fiske:Misner_Ylm}.
For now it suffices to follow Misner in denoting it as generically Hermitian.)
The inverse to this metric
$G^{AB}$ can be used to raise indices on functions defined on the sphere.

Making use of this new metric, and with some further analysis, the
approximation for the spherical harmonic coefficients
\begin{equation}
\Phi_{lm}(t,R) = \sum_{x\in\Gamma}\bar{R}_{lm}(x;R)w_{x}\Phi(t,x)
\label{eqn:PhiLMnum}
\end{equation}
follows with
\begin{equation}
R_{lm}(r;R) = \sum_{n=0}^{N} \bar{R}_{n}(R)Y^{nlm}(r,\theta,\phi)
\label{eqn:Rlm}
\end{equation}
in terms of $Y^{A} = G^{BA}Y_{B}$, not $Y_{A}$.

Note that one need only store the combination $w_{x}\bar{R}_{lm}$
at points where 
\(w_x \neq 0\) in order to compute the spherical harmonic components. 
This buries all the details of the computation in a relatively
small table of numbers, and, since these numbers are not time dependent, this
calculation need be done only once per simulation.

\section{Teukolsky Wave Solution}
\label{app:teuk}
The general form of the spacetime metric for the 
Teukolsky wave solution \cite{teuk:wave} is given by \eqnref{teuk}.
The angular function are
\begin{subequations} \label{eqn:angular}
\begin{eqnarray}
f_{rr} & = & 2 - 3\sin^{2}\theta \\
f_{r\theta} & = & -3\sin \theta \cos \theta \\
f_{r\phi} & = & 0 \\
f_{\theta\theta}^{(1)} & = & 3\sin^{2}\theta \\
f_{\theta\theta}^{(2)} & = & -1 \\
f_{\theta\phi} & = & 0 \\
f_{\phi\phi}^{(1)} & = & -f_{\theta\theta}^{(1)} \\
f_{\phi\phi}^{(2)} & = & 3\sin^{2}\theta - 1
\end{eqnarray}
\end{subequations}
for the $l=2$, $m=0$ case.
The remaining functions
\begin{subequations}
\label{eqn:teukABC}
\begin{eqnarray}
A & = & 3 \left( \frac{F^{(2)}}{r^3} + \frac{3F^{(1)}}{r^4}
	+ \frac{3F}{r^5} \right) \\
B & = & - \left( \frac{F^{(3)}}{r^2} + \frac{3F^{(2)}}{r^3}
	+ \frac{6F^{(1)}}{r^4} + \frac{6F}{r^5} \right) \\
C & = & \frac{1}{4} \left( \frac{F^{(4)}}{r} + \frac{2F^{(3)}}{r^2}
	+ \frac{9F^{(2)}}{r^3} + \frac{21F^{(1)}}{r^4}
	+ \frac{21F}{r^5} \right)
\end{eqnarray}
\end{subequations}
are written in terms of a free generating function $F=F(t - r)$, which
we choose to have the form given by \eqnref{F}.
The notation
\begin{equation}
F^{(n)} = \left[ \frac{d^{n}F(x)}{dx^n} \right]_{x = t - r}
\end{equation}
denotes various derivatives.  Taking $F$ as a function of $t-r$ corresponds
to outgoing waves.  To generate an ingoing solution, change the argument of $F$
to $t+r$, and change, in \eqnref{teukABC}, the sign in front of all of the 
terms with \emph{odd} numbers of 
derivatives. (Note that the description
of how ingoing waves are constructed in 
Ref.~\cite{goddard:waves} contains an error.
The description here, which matches the original reference, 
Ref.~\cite{teuk:wave}, is correct.)

The Weyl scalar $\Psi_4$ for this spacetime is computed from the 
definition \eqnref{Psi4} using the Kinnersley tetrad, 
\eqnref{Kinnersley},
and noting that, of the twelve non-zero components of the Riemann tensor
associated with the metric \eqnref{teuk}, only
\begin{subequations}
\begin{eqnarray}
R_{t\theta t\theta} & = &  
	-\frac{3}{2} r^{2} \sin^{2} \theta \fpderp{C}{t} 
	+ \frac{1}{2} r^{2} \fpderp{A}{t} \\
R_{t\phi t\phi} & = & 
	\frac{3}{2} r^{2} \sin^{4}\theta \left( \fpderp{C}{t}
	- \fpderp{A}{t} \right) \nonumber \\
& & \mbox{} + \frac{1}{2} r^{2} \sin^{2}\theta \fpderp{A}{t} \\
R_{t\theta r\theta} & = & 
	-\frac{1}{8} r^{3} \sin^{2} \theta \left(3 \fpderthree{B}{t}
	+ \fpderthree{A}{t} \right) \\
R_{r\phi r\phi} & = & -\sin^{2} \theta R_{t\theta t\theta} \\
R_{r\theta r\theta} & = & -\frac{1}{\sin^{2}\theta} R_{t\phi t\phi} \\
R_{t\phi r\phi} & = & -\sin^{2} \theta R_{t\theta r\theta}
\end{eqnarray}
\end{subequations}
contribute to the sum.

\bibliography{gr}

\end{document}